\documentclass[showpacs,preprintnumbers,preprint,aps]{revtex4}

\newcommand{\BE}{\begin{equation}}
\newcommand{\EE}{\end{equation}}
\newcommand{\BA}{\begin{eqnarray}}
\newcommand{\EA}{\end{eqnarray}}

\newcommand{\BW}{\begin{widetext}}
\newcommand{\EW}{\end{widetext}}
\input epsf.tex

\begin{document}
%\draft
%\preprint{0202}

%\twocolumn[\hsize\textwidth\columnwidth\hsize\csname@twocolumnfalse\endcsname

%\begin{center}

\title{Gravity waves over topographical bottoms: Comparison with the experiment}

\author{Meng-Jie Huang}\author{Chao-Hsien Kuo}
\author{Zhen Ye}\email{zhen@shaw.ca}
\affiliation{Wave Phenomena Lab, Department of Physics, National
Central University, Chungli, Taiwan}

\begin{abstract}

In this paper, the propagation of water surface waves over
one-dimensional periodic and random bottoms is investigated by the
transfer matrix method. For the periodic bottoms, the band
structure is calculated, and the results are compared to the
transmission results. When the bottoms are randomized, the
Anderson localization phenomenon is observed. The theory has been
applied to an existing experiment (Belzons, et al., J. Fluid Mech.
{\bf 186}, 530 (1988)). In general, the results are compared
favorably with the experimental observation.

\end{abstract}

\pacs{47.10.+g, 47.11.+j, 47.35.+i.}

\maketitle

%\newpage

\section{Introduction}

Multiple scattering occurs when waves propagate in media with
scatterers, leading to many interesting phenomena such as the
bandgaps in periodically structured media and the Anderson
localization in disordered media\cite{Anderson}. Within a bandgap,
waves are evanescent; when localized, they remain confined
spatially near the transmission point until dissipated. The
phenomenon of bandgaps and localization has been both extensively
and intensively studied for electronic, electromagnetic, and
acoustic systems, as summarized in Ref.~\cite{Sheng}.

Propagation and scattering of gravity waves over topographical
bottoms has also been and continues to be a subject of much
research. From practical perspectives, it is essential to many
important ocean engineering problems such as designing underwater
structures in minimizing the impact of water waves on banks. From
the point of fundamental research, the system of water waves may
also play an important role. Since water waves are a macroscopic
phenomenon, they may be monitored and recorded in a laboratory
scale. In this way, many significant phenomena of microscopic
scales may be demonstrated with water waves. This would be
particularly useful in facilitating the understanding of abstract
concepts which may have been initiated for quantum waves. A great
amount of papers and monographs has been published on water waves
over various topographical bottoms
\cite{Long,Bar,Miles,Peregrine,Meyer,Yue,JFM,JFM1,Evans,Newman,Papan,Mei}.
A comprehensive reference on the topic can be found in three
excellent textbooks \cite{Lamb,CCM,Ding}.

The concept of bandgaps and Anderson localization has been
extended to the the study of the propagation of water surface
waves over topographical bottoms. In 1983, Guazzelli {\it et al.}
\cite{Guazzelli} suggested that the phenomenon of Anderson
localization could be observed on shallow water waves, when the
bottom has random structures. Later, Devillard {\it et al.}
reconsidered water wave localization inside a channel with a
random bottom in a framework of the potential theory
\cite{Devillard}. They computed the localization length for
various cases. The experimental observation of water wave
localization has been subsequently suggested by Belzone {\it et
al.}\cite{Belzons}.

When the topographical bottoms are periodically structured, the
propagation of water surface waves will be modulated accordingly.
According to Bloch theorem, waves in a periodic medium, now termed
as Bloch waves, can be expressed in terms of the product of a
plane wave and a periodic function which is has the periodicity of
the medium. Therefore, the waves will exhibit the properties of
both plane wave propagation and periodic modulation. Indeed, a
recent experiment\cite{Nature} has used gravity waves to
illustrate the phenomenon of Bloch wave over a two dimensional
periodic bottom. This pioneering experiment has made it possible
that the abstract concept be presented in an unprecedentedly clear
manner. The experimental results have also been matched by a
theoretical analysis in Ref.~\cite{PRE}.

The experimental advantures\cite{Belzons,Nature} pave a new avenue
for investigating the phenomena of Anderson localization in
disordered media and wave bandgaps in periodic structures. In
particular, it has been recognized that the two dimension is a
marginal dimension, many peculiar wave phenomena could be
expected\cite{CMP}. Since gravity waves are naturally a two
dimensional phenomenon, it may be expected that water surface
waves can play a role in demonstrating some of the peculiar wave
phenomena.

Motivated by the developments, we wish to further consider the
propagation and localization properties of water surface waves. At
the first stage, we will consider water waves over one dimensional
uneven bottoms. We present a theoretical analysis of the previous
experimental results\cite{Belzons}. The formulation in
Ref.~\cite{Ye} will be used for the purpose. The comparison
between the experimental and theoretical results, in return,
provides a verification of the theory. We will study the band
structure of periodic cases, the effect of randomness on wave
propagation, the relation between the bandgaps and the
localization, and the amplitude or energy distribution over the
structured bottoms. The dependence on parameters, such as the
frequency, water depth, the variations of the height and width of
the obstacle steps, will be examined in detail. Although the
experiment to be analyzed here was done nearly twenty years ago,
to the best of our knowledge, however, there are no further
experiments which have been done on water waves in the context of
the localization effects. Even the existing limited experimental
results have not been thoroughly analyzed. The present paper is to
bridge the gaps, with the hope that further experimental
investigation may be arranged.

The paper will be constructed as follows. In the next section, we
will present the formulation and parametrization for the problem.
The results and discussion will be presented in section III,
followed by a summary in section IV.

\section{The general formulation}

A theory of water wave propagation over step-mounted bottoms has
been recently proposed and developed in Refs.~\cite{PRE,Ye}. This
formulation has been used earlier in interpreting some
experimental data \cite{PRE}. While the details can be referred to
in Ref.~\cite{Ye}, here we only present the final equation. After
the Fourier transformation, the equation describing the wave of
frequency $\omega$ over topographical bottoms is \BE
\nabla\left(\frac{1}{k^2}\nabla\eta\right) + \eta = 0,
\label{eq:finalb}\EE where $k$ satisfies \BE \omega^2 =
gk(\vec{r})\tanh(k(\vec{r})h(\vec{r})), \label{eq:dispersion}\EE
where $\eta$ is the surface displacement, $g$ is the gravity
acceleration constant, and $h$ is the depth from the surface. For
a fixed frequency, the variation of the wavenumber $k$ with the
topographical bottom is determined by the depth function $h$.

From Eq.~(\ref{eq:finalb}), we have the conditions linking domains
with different depths as follows: both $\eta$ and
$\frac{\tanh(kh)}{k}\eta = \frac{\omega^2}{gk^2}\eta$ are
continuous across the boundary.

\subsection{Application to one-dimensional situations}

\subsubsection{A single step}

\begin{figure}[hbt]
\epsfxsize=4in\epsffile{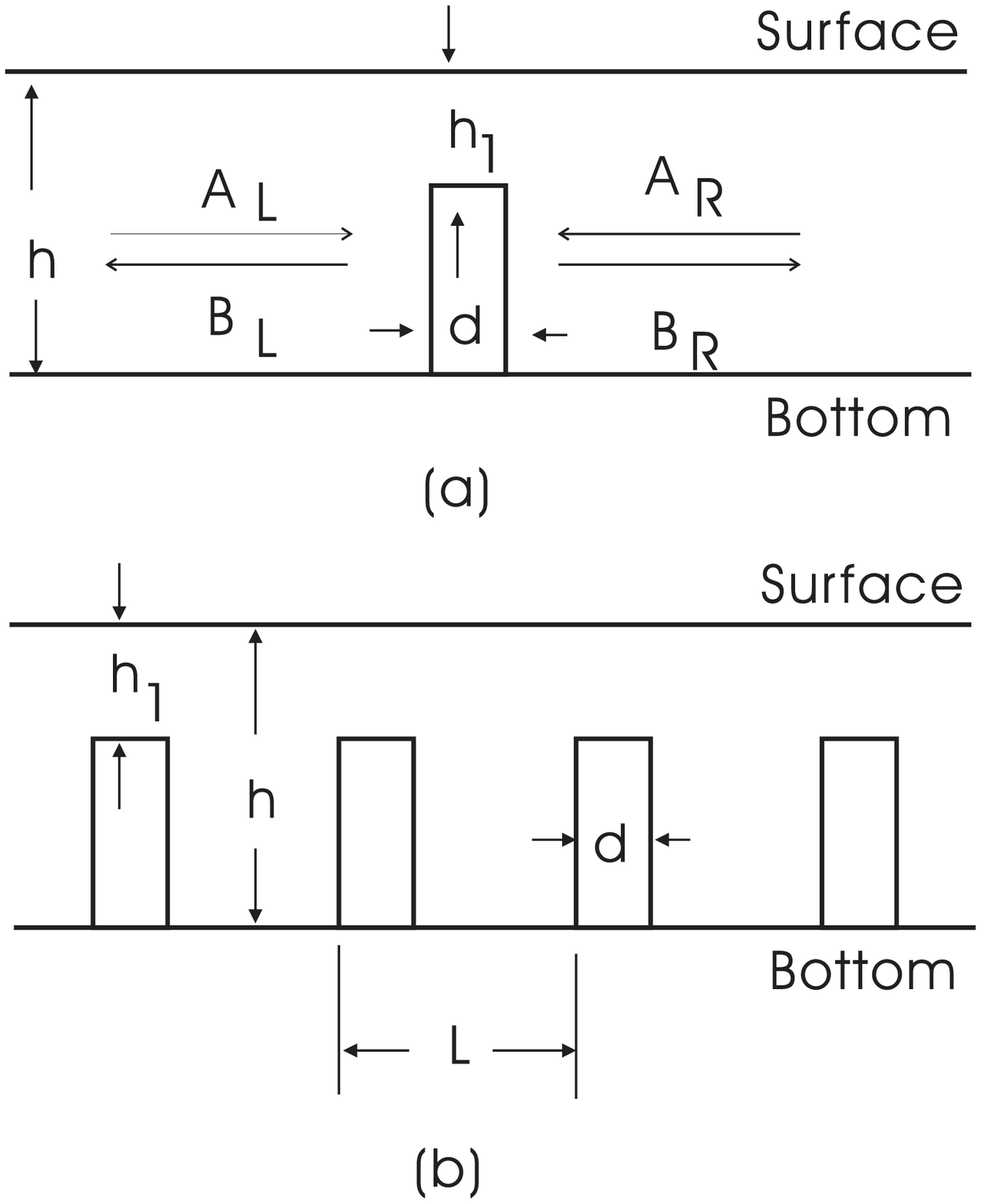} \smallskip \caption{Conceptual
layouts.} \label{fig1}
\end{figure}

First, consider a step with width $d$, and a wave is propagating
along the $x$ direction. The conceptual layout is in
Fig.~\ref{fig1}(a). We use the standard transfer matrix method to
solve for the wave transmission across the step.

The waves on the left, within, and on the right side of the step
can be generally rewritten as \BA \eta_L &=& A_L e^{ik_L x}
+ B_L e^{-ik_L x}\nonumber\\
\eta_M &=& A_M e^{ik_M x} + B_M e^{-ik_M x}\nonumber\\
\eta_R &=& A_R e^{ik_R x} + B_R e^{-ik_R x}. \EA The subscripts
$L, M, R$ represent the quantities on the left side, in the
middle, and on the right side of the step respectively.

The boundary conditions lead to the following equations: \BW\BA
A_L e^{ik_L x_L} + B_L e^{-ik_L x_L} &=& A_M e^{ik_M x_L} + B_M
e^{-ik_M
x_L}; \\
\frac{1}{k_L} \left( A_L e^{ik_L x_L} - B_M e^{-ik_L x_L} \right)
&=& \frac{1}{k_M} \left(A_M e^{ik_M x_L} - B_M e^{-ik_M
x_L}\right), \EA \EW and \BW\BA A_M e^{ik_M x_R} + B_M e^{-ik_M
x_R} &=& A_R e^{ik_R x_R} + B_R e^{-ik_R x_R};\\ \frac{1}{k_M}
\left(A_M e^{ik_M x_R} - B_M e^{-ik_M x_R} \right) &=&
\frac{1}{k_R} \left(A_R e^{ik_R x_R} - B_R e^{-ik_R x_R}\right).
\EA\EW In these equations, $x_{L,R}$ stands for the locations of
the left and right sides of the step respectively, and $x_R - x_L
= d$.

The first set of boundary equations gives the matrix relation \BE
\left(
\begin{array}{c} A_L \\ B_L \end{array}\right) = T_{LM} \left(
\begin{array}{c}A_M \\ B_M
\end{array}\right)\EE with \BW
\BE T_{LM} = \frac{1}{2} \left( \begin{array}{cc} (1+g_{LM})
e^{i(k_M-k_L) x_L} & (1-g_{LM}) e^{-i(k_M + k_L)x_L} \\
(1-g_{LM})e^{i(k_M + k_L)x_L} & (1+g_{LM}) e^{-i(k_M - k_L) x_L}
\end{array} \right), \label{eq:LM}
\EE \EW and $$ g_{LM} = \frac{k_L}{k_M}.$$

Similarly, we can derive \BE \left(
\begin{array}{c} A_M \\ B_M \end{array}\right) = T_{MR} \left(
\begin{array}{c}A_R \\ B_R
\end{array}\right)\EE with \BW
\BE T_{MR} = \frac{1}{2} \left( \begin{array}{cc} (1+g_{MR})
e^{i(k_R-k_M) x_R} & (1-g_{MR}) e^{-i(k_R + k_M)x_R} \\
(1-g_{MR})e^{i(k_R + k_M)x_R} & (1+g_{MR}) e^{-i(k_R - k_M) x_R}
\end{array} \right)\label{eq:MR} \EE \EW and $$ g_{MR} =
\frac{k_M}{k_R}.$$

From Eqs.~(\ref{eq:LM}) and (\ref{eq:MR}), we obtain the following
solution in the transfer matrix form \BE \left(\begin{array}{c} A_L \\
B_L \end{array}\right) = T_{LR}\left(
\begin{array}{c} A_R \\ B_R \end{array}\right) \label{eq:LR}
\EE with \BE T_{LR} = T_{LM}T_{MR}.\EE Eq.~(\ref{eq:LR}) relates
the waves on the left to the right side of the step.

\subsubsection{The case of $N$ steps}

Now we consider $N$ steps in a unform medium of wave number $k$.
The illustration is in Fig.~\ref{fig1}(b). The step widths are
$d_i$ and the water depths over the steps are $h_i$. The wave
number over the step is denoted by $k_i$ ($i=1,\dots, N$). They
satisfy the following relations respectively, \BE \omega^2 =
gh\tanh(kh), \ \ \ \omega^2 = gh_i\tanh(k_ih_i).\EE

Clearly, the coefficients on the most left hand region is related
to the most right hand region by \BE
\left(\begin{array}{c} A_L \\
B_L \end{array}\right) = T(N)\left(
\begin{array}{c} A_R \\ B_R \end{array}\right) \label{eq:final}
\EE with \BE T(N) = \prod_{i=1}^{N} T_i\EE The matrix $T_i$ is the
transfer matrix for the $i$-th step and will be given below.

Let us consider a unit plane wave propagation along the $x$
direction, and explore the reflection and transmission properties.
In this case, clearly we have \BE A_L = 1, \ \ \ B_R = 0.\EE $B_R
= 0$ is the common radiation condition. Thus from
Eq.~(\ref{eq:final}) we arrive at the solutions \BE A_R(N) =
\frac{1}{T_{11}(N)}, \ \ \ B_L(N) = \frac{T_{21}(N)}{T_{11}(N)}.
\label{eq:AB}\EE The subscripts denote the corresponding matrix
elements.

The transmission and reflection coefficients are defined as \BE T
= |A_R(N)|^2, \ \ \ R = 1 - T.\EE

Now we construct the $T$ matrix for each step. In the current
case, we have \BE g_{LM}(i) = \frac{k}{k_i}, \ \ g_{MR}(i) =
\frac{k_i}{k},\EE and \BE k_L = k, \  k_M = k_i, \ k_R = k. \EE
 We denote $g_{s,i} =
\frac{k}{k_i}$. Therefore \BW \BA T_i &=& \frac{1}{4} \left(
\begin{array}{cc} (1+g_{s,i})
e^{i(k_i-k) x_{i,L}} & (1-g_{s,i}) e^{-i(k_i + k)x_{i,L}} \\
(1-g_{s,i})e^{i(k + k_i)x_{i,L}} & (1+g_{s,i}) e^{-i(k_i - k)
x_{i,L}}
\end{array} \right)\nonumber\\ & & \times \left( \begin{array}{cc} (1+1/g_{s,i})
e^{i(k-k_i) x_{i,R}} & (1-1/g_{s,i}) e^{-i(k + k_i)x_{i,R}} \\
(1-1/g_{s,i})e^{i(k + k_i)x_{i,R}} & (1+1/g_{s,i}) e^{-i(k - k_i)
x_{i,R}}
\end{array} \right) \label{eq:trans}
\EA \EW

\subsection{Simulation setup}

\subsubsection{Non-dimensional parametrization}

Consider an infinite periodic array of the steps, as shown in
Fig.~\ref{fig1}. The lattice constant is $L$. For random arrays,
$L$ refers to the average distance between two adjacent steps.

The dispersion relation is
$$\omega^2 = gk\tanh(kh).$$ This can be rewritten as
$$\frac{\omega^2}{\omega^2_0} =
(kL)\tanh\left((kL)\frac{h}{L}\right),$$ with $$\omega_0^2 =
\frac{g}{L}.$$ Therefore in all later computations, the length can
be scaled by $L$, the frequency by $\omega_0$, the wavenumber by
$kL$.

The wavenumbers in the medium and within the steps are given by
(at the same frequency) \BA \frac{\omega^2}{\omega^2_0} &=&
(kL)\tanh\left((kL)\frac{h}{L}\right),
\\
\frac{\omega^2}{\omega^2_0} &=&
(k_iL)\tanh\left((k_iL)\frac{h_i}{L}\right).\EA This leads to
$$g_{s,i} = \frac{kL}{k_iL},$$ and the transfer matrix of the
$i$-th step is \BW \BA T(i) &=& \frac{1}{4} \left(
\begin{array}{cc} (1+g_{s,i})
e^{i(k_iL-kL) \frac{x_{i,L}}{L}} & (1-g_{s,i}) e^{-i(k_iL + kL)\frac{x_{i,L}}{L}} \\
(1-g_{s,i})e^{i(kL + k_iL)\frac{x_{i,L}}{L}} & (1+g_{s,i})
e^{-i(k_iL - kL) \frac{x_{i,L}}{L}}
\end{array} \right)\nonumber \\ & &\times \left( \begin{array}{cc} (1+\frac{1}{g_{s,i}})
e^{i(kL-k_iL) \frac{x_{i,L} + d_i}{L}} & (1-\frac{1}{g_{s,i}}) e^{-i(kL + k_iL)\frac{x_{i,L}+d_i}{L}} \\
(1-\frac{1}{g_{s,i}})e^{i(kL + k_iL)\frac{x_{1,L}+d_i}{L}} &
(1+\frac{1}{g_{s,i}}) e^{-i(kL - k_iL) \frac{x_{i,L}+d_i}{L}}
\end{array} \right),%\nonumber%\label{eq:transm}
\EA\EW where $x_{i,L}$ is the coordinate of the left side of the
$i$-th step.

\subsubsection{Band structure for periodic cases}

For the periodically arranged steps with $d_i=d$, and $h_i=h_1$,
the band structure can be solved. According to Bloch theorem, the
water surface displacement $\eta$ can be expressed as \BE \eta(x)
= \xi(x)e^{iKx}, \label{eq:B}\EE where $K$ is the Bloch
wavenumber, and $\xi(x)$ is a periodic function modulated by the
periodicity of the structure, i.~e. $\xi(x+L) = \xi(x).$ The
relation between $K$ and the frequency $\omega$ can be obtained by
taking Eq.~(\ref{eq:B}) into Eq.~(\ref{eq:final}).

We can derive an equation determining the band structure in the
periodic case, \BW\BE \cos (KL) = \cos (k_1L (d/L)\cos(kL(L-d)/L)
- \cosh(2\xi)\sin(k_1L(d/L))\sin(kL(L-d)/L),\EE \EW where $$\xi =
\ln(q), \ \ \ \mbox{with} \ \ \ \omega^2 = gk_1\tanh(kh_1),\ \ \
q^2 = \frac{k_1}{k}.$$

\subsubsection{Random situations}

There are a number of ways to introduce the randomness. (1)
Variation in the height of the steps: With the fixed widths and
positions of the steps, the height of the steps can be varied in a
controlled way. For example, the height of the steps can be varied
randomly between $[H_0 - \Delta H, H_0 + \Delta H]$. (2)
Positional disorders: Initially, the steps can be arranged in a
lattice form. Then allow each step to move randomly around its
initial position. The allowing range for movement can be
controlled and denotes the level of randomness. The extreme case
is completely randomness. (3) Width randomness: We can also
introduce the randomness for the widths of the steps. In the
simulation, we will consider the randomness introduced in the
experiment\cite{Belzons}.

When randomness is introduced, a few definitions are in order. The
most important quantity is the Lyapounov exponent $\gamma$. Its
definition is \BE \gamma = \lim_{N\rightarrow \infty} \langle
\gamma_N\rangle, \label{eq:gammal}\EE where $$ \gamma_N \equiv
-\frac{1}{N}\ln(|A_R(N)|).$$ Here $|A_R(N)|^2$ is the transmission
coefficient for a system with $N$ random steps, referring to
Eq.~(\ref{eq:AB}),
$$|A_R(N)|^2 = \frac{1}{|T_{11}(N)|^2}.$$ The symbol $\langle
\cdot\rangle$ denotes the average over the random configuration.
The inverse of the Lyapounov exponent characterizes the
localization length, i.~e. $\xi = \gamma_N^{-1}$.

\section{The results and discussion}

\begin{figure}[hbt]
\epsfxsize=3in\epsffile{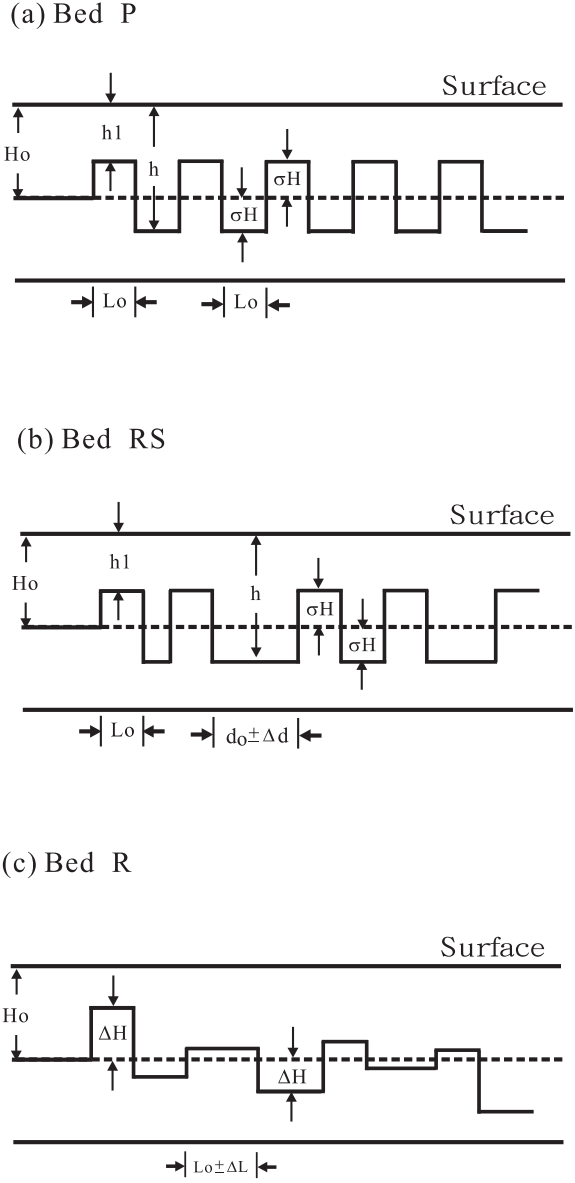} \smallskip \caption{Situations
considered in the paper, adopted from Fig.~2 of
Ref.~\cite{Belzons}. (a) Bed P case: in this case, the steps are
mounted periodically with lattice constant $2L_0$; the variation
of the steps $\sigma H$ is fixed; (b) Bed RS case: in the case,
the steps are allowed to move randomly from their initial periodic
positions, as set in the Bed P case - the allowed range is denoted
as $\pm \Delta L$ and the variation of the steps $\sigma H$ is
fixed; (c) Bed R: in this case, both the heights and the widths
are allowed to vary randomly from their initial values in the Bed
P case within the ranges [$H_0-\Delta H$, $H_0 +\Delta H$], and
[$L_0-\Delta L_0$, $L_0+\Delta L_0$].} \label{fig2}
\end{figure}

The systems are from the previous experiment\cite{Belzons}. That
is, the bottoms are mounted with a series of steps and these steps
are either regularly or randomly but on-average regularly placed
on the bottoms. Three cases are considered and are illustrated by
Fig.~\ref{fig2}. In the Bed P case, the averaged water depth is
$H_0$, the periodicity is $2L_0$, and the step variation is fixed
at $\sigma H$. In the Bed RS case, the averaged water depth is
$H_0$, the step variation is fixed at $\sigma H$, the separation
between steps is uniformly distributed with $[d_0 -\Delta d, d_0 +
\Delta d]$. In the Bed R case, both the height and the separation
between the steps are allowed to vary randomly, but within the
ranges $[H_0 - \Delta H, H_0 + \Delta H]$ and $[L_0 -\Delta L, L_0
+ \Delta L]$ respectively.

The experimental setups have been described in Section 2 of
Ref.~\cite{Belzons}. We briefly repeat here. The experiments were
carried out in a glass-walled wave tank with length 4m and width
0.39m. A bottom composed of periodic or random steps was built
into a flat bottom with the mean water depth $H_0$. The different
bottoms varied only along the tank so that, apart from weak edge
effects, the propagation of waves is considered to be
one-dimensional. The resolution of the water depth is estimated at
about 0.2mm.

\subsection{Band structure and transmission}

\begin{figure}[hbt]
\epsfxsize=5in\epsffile{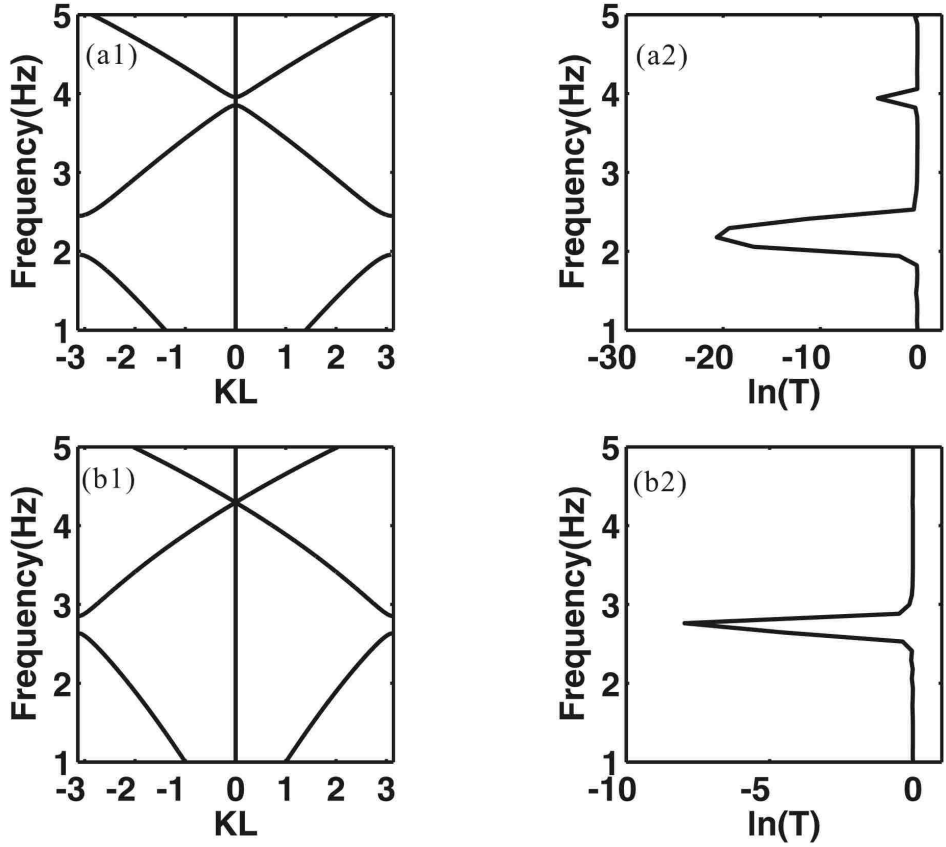} \smallskip \caption{Band
structure and transmission for the Bed P case in two situations,
referring to Fig.~\ref{fig2}(a): (a1) and (a2) the average water
depth is $H_0 = 1.75$cm, the height variation is $\sigma H/H_0 =
0.43$; (b1) and (b2) the average water depth is $H_0 = 3$cm, the
height variation is $\sigma H/H_0 = 0.25$. The left and right
panel show the band structure and transmission results
respectively. The transmission is presented in the log scale for
100 steps.} \label{fig3}
\end{figure}

First, we consider the first case in the experiment: the periodic
case, i.~e. the Bed P case. For this case, the band structure and
the transmission are computed for two water depths. In both cases,
the width of the steps is $L_0 = 4.1$cm; therefore the periodicity
is 8.2cm. The results are shown in Fig.~\ref{fig3}. From the band
structure results in (a1) and (b1), we observe that for the small
water depth ($H_0=1.75$cm), there are two bandgaps in the
frequency range measured in the experiment, while in the deeper
case ($H_0=3$cm) there is one bandgap. The locations of the band
gaps match the inhibited transmission regimes. The width of the
gap and the inhibition effect tend to decrease with frequency as
shown by (a1) and (a2). This is understandable. In the high
frequency limit, i.~e. when $kh >> 1$, the dispersion relation in
Eq.~(\ref{eq:dispersion}) approaches $\omega^2 = gk.$ Therefore
the importance of the bottom structure will decrease with
increasing frequency. The results in Fig.~\ref{fig3} will help us
comprehend the later results.

\subsection{Reflection coefficient}

\begin{figure}[hbt]
\epsfxsize=6in\epsffile{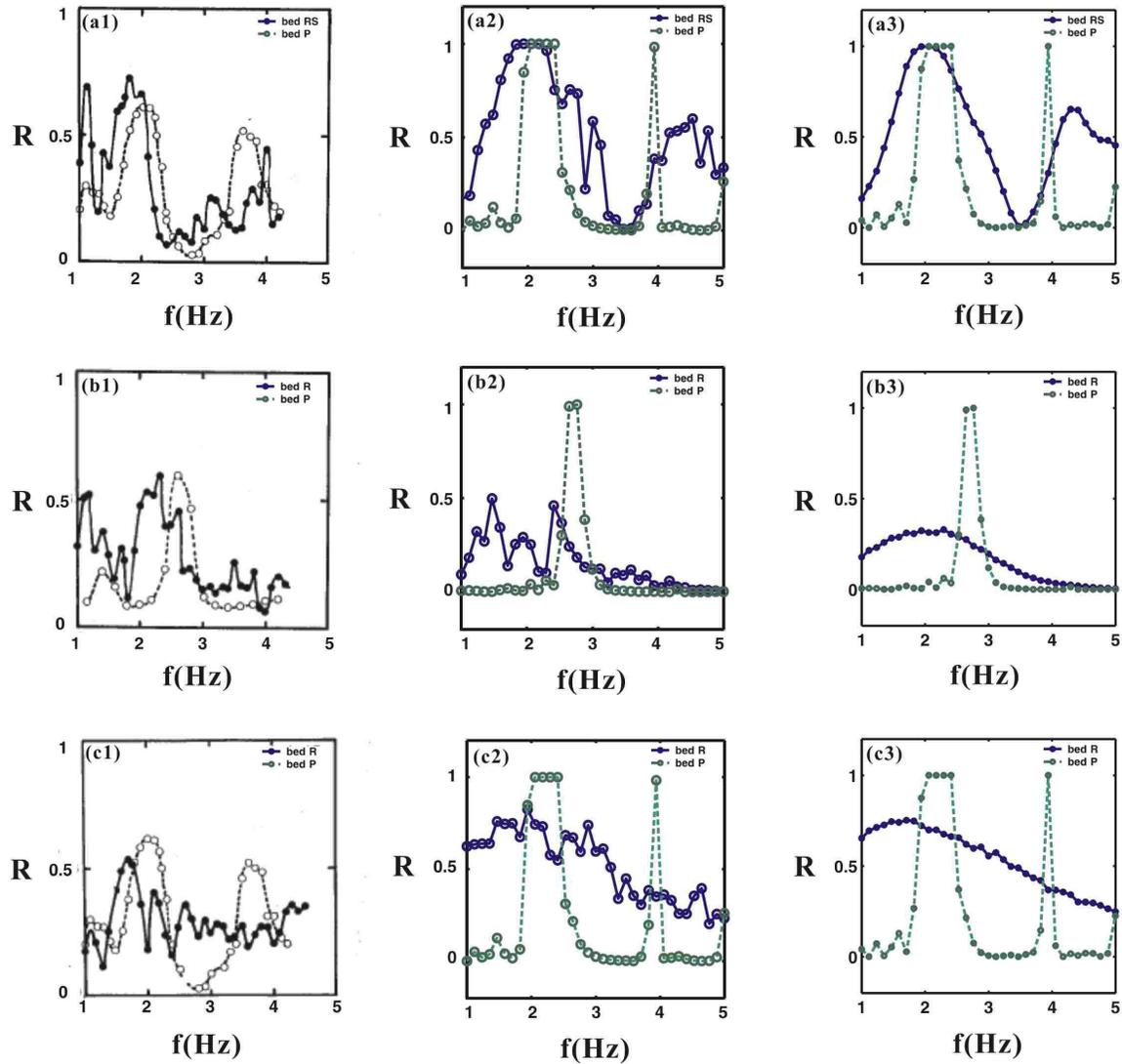} \smallskip \caption{Reflection
versus frequency for the Bed RS and P cases with three average
water depths. Left panel: The results from the
experiment\cite{Belzons}. Middle panel: the theoretical results
with the average over five random configurations. Right panel: the
theoretical results with the average over ten thousand random
configurations, so to make sure the stability of the average.}
\label{fig4}
\end{figure}

In the experiment\cite{Belzons}, the reflection coefficients are
measured for the three cases: Bed P, Bed R and Bed RS cases. Two
average water depths are considered: $H_0 = 1.75$ and 3 cm. We
have considered all the cases, and applied the formulation in
Section II to obtain corresponding results. In Fig.~\ref{fig4}, we
present our theoretical results. For the convenience of the reader
and as a comparison, we also re-plot the experimental results on
the same figure (left panel). We have take into account two random
configuration numbers in the simulation: one is 5 random
configuration, i.~e. the middle panel, which is take as the same
as in the experiment; the other in the right panel is more than
ten thousand random configurations to ensure the stability of the
averaging results. All the parameters are repeated from
Ref.~\cite{Belzons}.

Fig.~\ref{fig4}(a1), (a2) and (a3) compare the results for the Bed
RS and Bed P cases with averaged water depth $H_0 = 1.75$ and step
width $L_0 = 4.1$cm. For both cases, the ratio $\sigma H/H_0$ is
fixed at 0.43, i.~e. the there is no variation in the step
heights. In the Bed RS case, the disorder is introduced to the
separation between steps, that is, the separation is randomly
chosen with uniform distribution within [2cm, 8cm] or [$d_0-3$,
$d_0+3$] with $d_0=5$cm. In the simulation, the total number of
steps is 58. We have taken two numbers of random configuration in
the simulation. One is five (a2), which complies with the
experiment, and the other (a3) is ten thousand time, so to ensure
the stability of the averaging. The experimental data are shown in
(a1). The comparison of (a1), (a2) and (a3) indicates the
following. Overall speaking, the theoretical results capture well
the qualitative features observed experimentally, and agree to
certain extend with the experimental results.

First we consider the Bed P case. (1) The theoretically predicted
positions of the reflection peaks agree well with the experimental
observation in the Bed P case. These positions also coincide the
bandgaps from the band structure computation in Fig.~\ref{fig1}.
(2) In the Bed P case, the reflection coefficient reaches its
maximum value of one within the bandgaps as expected, while the
experimental values are always smaller than one for the frequency
range considered. A possible reason for this discrepancy may be
that in the theoretical simulation, we did not take into account
possible dissipation effects caused by such as viscosity and
thermal exchange; some effects have been discussed in
Ref.~\cite{Belzons}. (3) The theoretical width of the first
reflection peak in the Bed P case matches well that observed, but
the theoretical width of the second reflection peak at about 4 Hz
is narrower than that from the experiment. In fact, the
experimentally measured widths of the two reflection peaks are
more or less the same. Since the effects of the periodic bottom
diminish with increasing frequency as discussed above, we may
conclude that there are other effects which could broaden the
reflection peak at 4 Hz, and these effects may include those from
the dissipation, non-linearity, and evanescent mode leakage. These
effects have not been considered in the current theory\cite{Ye}.

Now we consider the Bed RS case. (1) Again, except for the peak
values in the reflection, the theoretically results reproduce the
experimental observation reasonably well in general, particularly
at the strong reflection located at about 2Hz. (2) Different from
the Bed P case, the width of the first reflection peak at 2Hz is
wider in theory than in the experiment. (3) In both the theory and
experiment, a second reflection peak is noticed within 4 and 5 Hz.
(4) An obvious difference between the theory and the experiment is
at the low frequency around 1.2Hz: A strong sharp reflection peak
appears at 1.2Hz in the experiment, but absent in the theory. This
experimental observation differs from previous observation in
acoustic or optical systems\cite{Sheng}. In acoustic or optical
systems, the disorder effect decreases as the frequency decreases.
Therefore, waves tend to diffuse away, leading to weaker
reflections at low frequencies. In fact, the result of the
reflection measurement is also in disagreement with the
localization measurement shown in Fig.~15 of Ref.~\cite{Belzons}
where it is shown that the localization length at 1.2Hz is even
longer than at 1.5Hz at which the reflection is small; the longer
the localization length, the weaker is the reflection. (5)
Increasing the number of random configuration tend to smooth the
curves.

The comparison between the theoretical and experimental reflection
results for the Bed P and Bed R cases with $H_0 = 3$cm is shown by
Fig.~\ref{fig4}(b1), (b2) and (b3). The parameters are as follows.
(1) Bed P: $\sigma H/H_0$ = 0.25, $L_0=4.1$cm; (2) Bed R: the
separation between steps varies completely randomly within
$L_0\pm\Delta L$ with $\Delta L = 2$cm, and the height of the
steps varies uniformly within $H_0\pm\Delta H$ with $\Delta H =
1.26$cm. The number of steps is 58. In (b2), five random
configurations are used for averaging, and in (b3) ten thousand
random configurations are used to ensure the stability of the
averaging. In the Bed P case, except at the reflection peak, the
theoretical results reproduce very well the experimental
observation. In the Bed R case, the theoretical results also match
that from the experiment in both the qualitative structure and the
magnitude, referring to (b1) and (b2). The existing deviation may
result from the insufficient random average.

The comparison between the theoretical and experimental reflection
results for the Bed P and Bed R cases with $H_0 = 1.75$cm is shown
by Fig.~\ref{fig4}(c1), (c2) and (c3). The parameters are as
follows. (1) Bed P: $\sigma H/H_0$ = 0.43, $L_0=4.1$cm; (2) Bed R:
the separation between steps varies completely randomly within
$L_0\pm\Delta L$ with $\Delta L = 2$cm, and the height of the
steps varies uniformly within $H_0\pm\Delta H$ with $\Delta H =
1.24$cm. The number of steps is 58. In (c2), five random
configurations are used for averaging, and in (c3) ten thousand
random configurations are used to ensure the stability of the
averaging. The Bed P case has been discussed in the above. In the
Bed R case, the general features of the experimental and
theoretical results seem to be agreeable with each other. The
predicted reflection curve starts to match qualitatively the
experimental data from about 3 Hz. The discrepancy at low
frequencies is, again, noticeable.

\subsection{Localization length}

In the experiment\cite{Belzons}, the localization length is
extracted from the measurement of the total wave amplitude
attenuations. In the simulation, the localization length is
obtained from the inverse of the Lyapounov exponent given in
Eq.~(\ref{eq:gammal}). Here the Bed R case is considered and the
parameters are: $H_0 = 1.75$cm, $L_0=4.1$cm, and the height of the
steps and the separation between steps vary randomly within the
ranges [$H_0-\Delta H, H_0 + \Delta H$] and [$L_0 - \Delta L,
L_0+\Delta L$] respectively; here $\Delta H = 1.2425$cm, and
$\Delta L = 2$cm. Ten thousand steps and ten thousand random
configurations have been used in the simulation to ensure the
stability of the numerical results.

\begin{figure}[hbt]
\epsfxsize=6in\epsffile{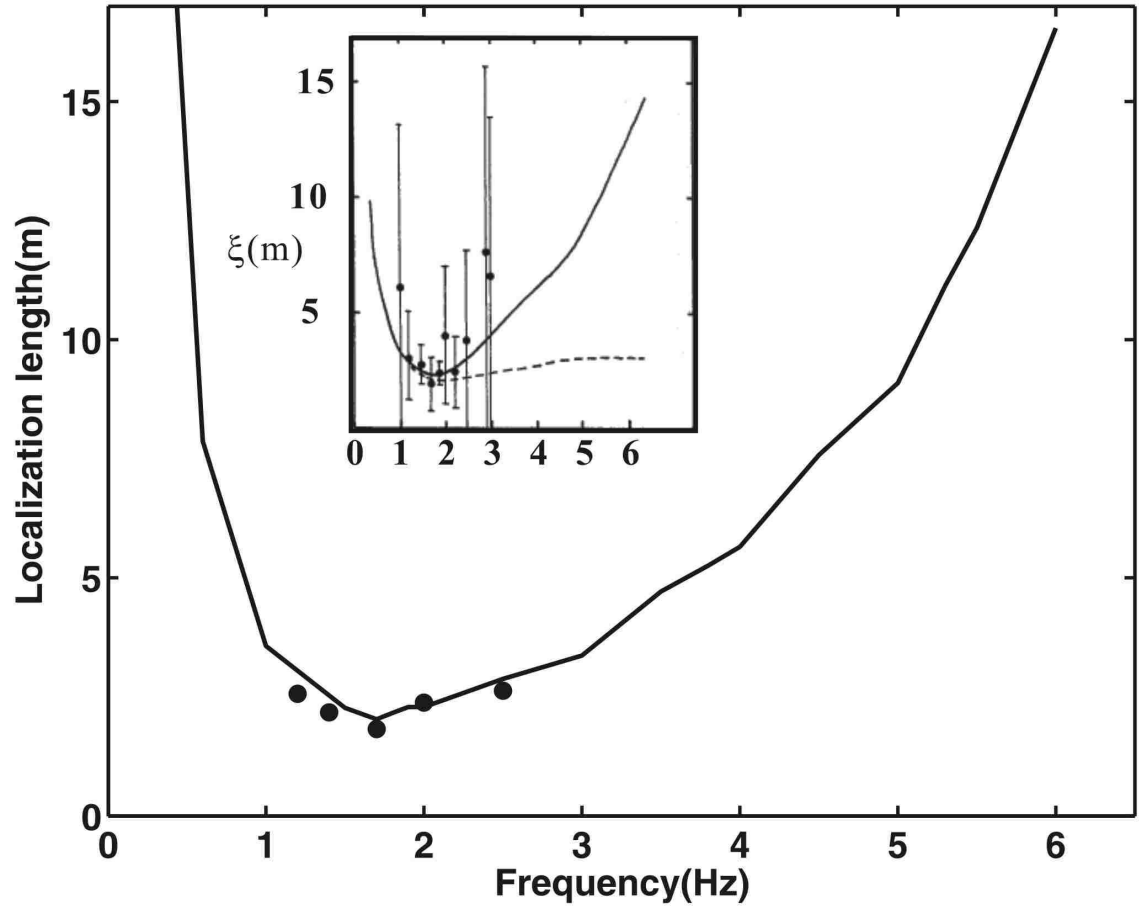} \smallskip \caption{Localization
length versus frequency for the Bed R case. The experimental
results are shown in the inset figure. The five black dots denote
the results from an averaging over five random configurations.}
\label{fig5}
\end{figure}

The numerical and experimental results are shown in
Fig.~\ref{fig5}. Here the localization length is plotted against
the frequency. The results from Ref.~\cite{Belzons} are shown in
the inserted box. A few observations are in order. (1) In
Ref.~\cite{Belzons}, the authors have used a potential formulation
to obtain the localization length, denoted by the solid length in
the inserted box. By eye-inspection, we see that the present
numerical results agree remarkably well with the results from the
potential theory, thus providing another support for the present
relatively simple theory, stemmed from Ref.~\cite{Ye}. (2) The
numerical results also agree with the averaged experimental data
in the vicinity of the frequency 2Hz. (3) There is a huge
fluctuation in the experiment results. From our simulation, we
think that such a significant deviation is due to the insufficient
average numbers, an obvious limitation on any experiment. This is
particularly an important factor when the localization length is
long. Nevertheless, the agreements shown in Fig.~\ref{fig5} is
encouraging.

\subsection{Behavior of the wave amplitude along the random bed}

In the experiment, the variation of the wave amplitude along the
random bed is also measured. Both Bed RS and Bed R cases are
considered. The parameters used in the experiment\cite{Belzons}
are summarized as follows. In the Bed RS case, $H_0 = 1.75$cm,
$\sigma H/H = 0.43$, $L_0=4.1$cm, and the separation between steps
varies randomly in the range of [2cm, 8cm]. In the Bed R case,
$H_0 = 1.75$cm, $L_0 = 4.1$cm, and the height of the steps and the
separation between steps vary randomly within the ranges
[$H_0-\Delta H, H_0 + \Delta H$] and [$L_0 - \Delta L, L_0+\Delta
L$] respectively; here $\Delta H = 1.2425$cm, and $\Delta L =
2$cm. Four different frequencies have been measured and simulated.

\begin{figure}[hbt]
\epsfxsize=6in\epsffile{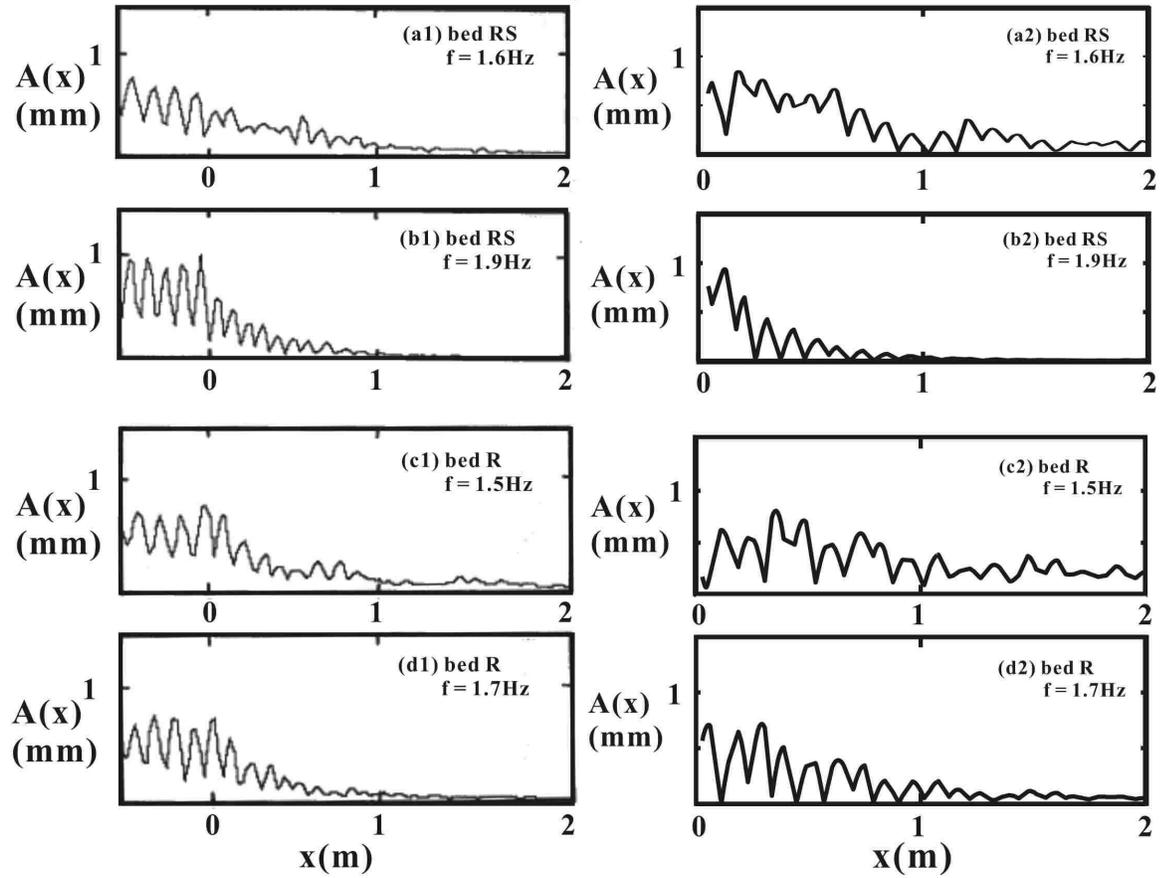} \smallskip \caption{Variation of
the amplitude of wave elevation along the wave tank for the Bed RS
and Bed R cases for different frequencies. The experimental
\cite{Belzons} and numerical results are shown on the left and
right panels respectively.} \label{fig6}
\end{figure}

The experimental and simulation results for a given random
realization of the random beds are presented in Fig.~\ref{fig6}.
It is shown that the theoretical results match remarkably well the
experimental results. It is shown that the waves do not decay
monotonically along the random bottom (without averaging), due to
the manifestation of resonant modes of the beds. The resonances
are sensitive to the frequency variation. We also found that the
occurrence of the resonances are sensitive to the random
configuration.

\begin{figure}[hbt]
\epsfxsize=6in\epsffile{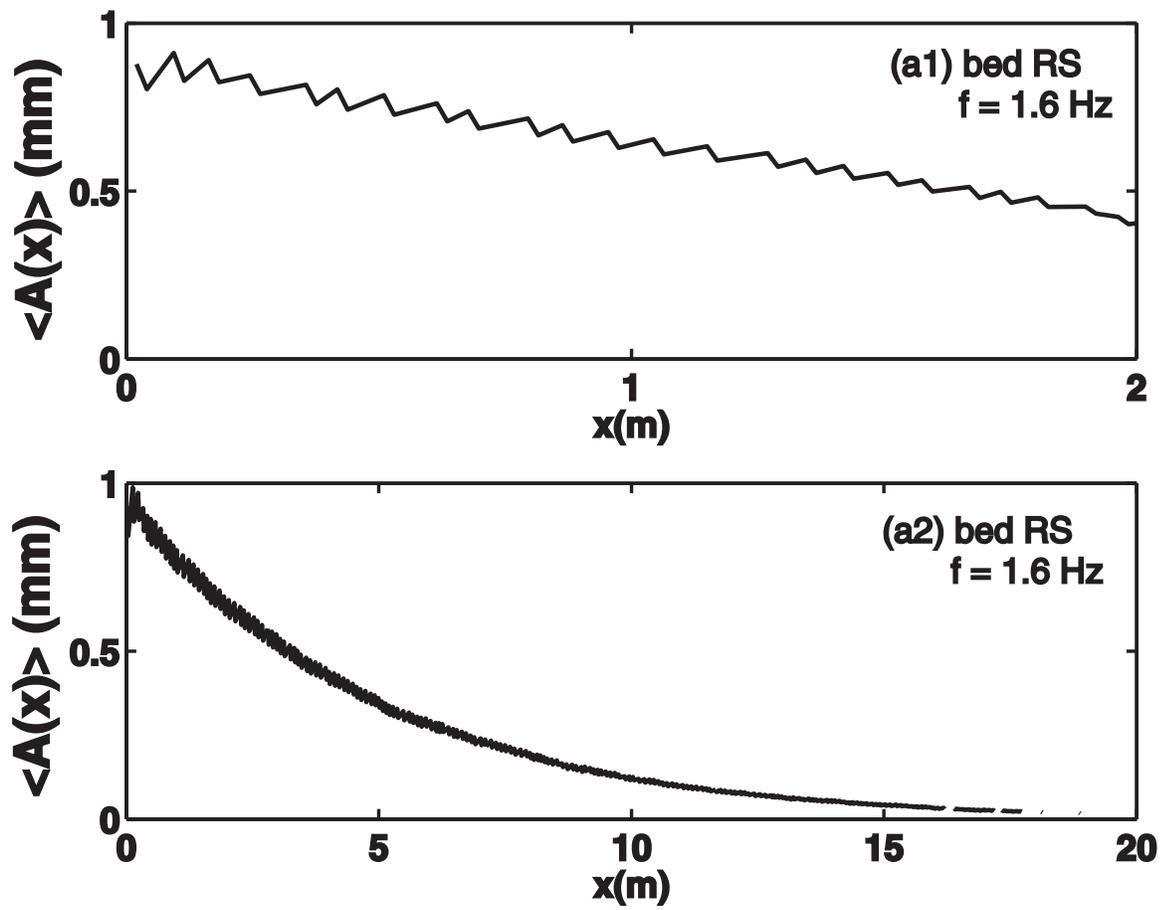} \smallskip \caption{The averaged
variation of the amplitude of wave elevation along the wave tank
for the Bed RS from Fig.~\ref{fig6} with $f = 1.6$Hz. To show the
behavior near the transmission site, the results are plotted in
two length scales: (a1) up to 2m; and (a2} up to 20m.
 \label{fig7}
\end{figure}

We have further computed the averaged variation of the wave
amplitude along the random bed for a sufficiently large number of
random configurations. We found that though smeared out a little
by the averaging, the resonance feature remains for spatial points
near the transmission and tends to diminish for large travelling
paths. And the averaged amplitude decays exponentially with
increasing travelling distances. As an example, in Fig.~\ref{fig7}
we illustrate these by the results of the Bed RS case with f =
1.6Hz. The results in Fig.~\ref{fig7} also indicate that the
exponential decay rate, associated with the localization length,
may not be accurately obtained from measurements done on
insufficiently long samples, as the fluctuation can be quite
significant for small sample sizes.

\section{Summary}

In summary, we have considered the propagation of water surface
waves over topographical bottoms. A transfer method has been
developed to compute the wave field along the propagating path,
the transmission and reflection coefficients. The localization
effects due to disordered bottom structures are also considered.
The theory has been applied to analyze the existing experimental
results. Some agreements and discrepancies are discovered and
discussed. It is pointed out that more detailed experiments may be
helpful in not only identifying the peculiar localization
phenomenon, but in helping improve theories for water wave
propagation over rough bottoms.

\acknowledgments

The work received support from the National Science Council
(NSC-92-2611-M008-002).

\newpage

\newpage

\newpage

\newpage

\newpage

\newpage

\newpage

\end{document}